\documentclass[journal]{IEEEtran}

\ifCLASSINFOpdf

\else

\fi

\usepackage{amsthm}
\usepackage{epsfig} 
\usepackage{epstopdf}
\usepackage{amsmath}
\usepackage{amsthm}
\usepackage{comment}
\usepackage{url}
\usepackage{algorithm}
\usepackage{algorithmic}

\usepackage{amssymb}
\usepackage{amsmath}
\usepackage{enumerate}
\usepackage{autobreak}
\usepackage{multirow} 
\usepackage{makecell}	
\usepackage{threeparttable} 
\usepackage{amsfonts}
\usepackage{bm}
\usepackage{color}
\usepackage{cite}
\usepackage{stfloats}
\usepackage{chemarrow}
\usepackage{extarrows}
\usepackage{enumitem} 
\allowdisplaybreaks

\theoremstyle{definition}
\newtheorem{theorem}{Theorem}

\newtheorem{proposition}[theorem]{Proposition}

\theoremstyle{definition}
\newtheorem{definition}{Definition}
\newtheorem{remark}{Remark}

\newtheorem{assumption}{Assumption}

\usepackage{comment}
\usepackage{ifthen}
\newboolean{showcomments}
\setboolean{showcomments}{true}
\newcommand{\su}[1]{\ifthenelse{\boolean{showcomments}}
	{ \textcolor[rgb]{1,0,1}{(ZW:  #1)}}{}}
\newcommand{\fliu}[1]{\ifthenelse{\boolean{showcomments}}
	{ \textcolor{red}{(FL:  #1)}}{}}
\newcommand{\slow}[1]{\ifthenelse{\boolean{showcomments}}
	{ \textcolor{blue}{(SL:  #1)}}{}}

\title{Sharing Energy in Wide Area: A Two-Layer Energy Sharing Scheme for Massive Prosumers}

\begin{document}
\graphicspath{{Figures/}}
\newcounter{MYtempeqncnt}

\author{
	Yifan Su,
	Peng Yang,
	Kai Kang,
	Zhaojian Wang, \IEEEmembership{Member, IEEE}, \\
	Ning Qi, \IEEEmembership{Member, IEEE},
	Tonghua Liu,
	and Feng Liu, \IEEEmembership{Senior Member, IEEE}


%
%
%
%
%

%
}

\markboth{Journal of \LaTeX\ Class Files,~Vol.~xx, No.~xx, xx~xxxx}%
{Shell \MakeLowercase{\textit{et al.}}: Bare Demo of I EEEtran.cls for IEEE Journals}

\maketitle

\begin{abstract}
The popularization of distributed energy resources transforms end-users from consumers into prosumers. Inspired by the sharing economy principle, energy sharing markets for prosumers are proposed to facilitate the utilization of renewable energy. This paper proposes a novel two-layer energy sharing market for massive prosumers, which can promote social efficiency by wider-area sharing. In this market, there is an upper-level wide-area market (WAM) in the distribution system and numerous lower-level local-area markets (LAMs) in communities. Prosumers in the same community share energy with each other in the LAM, which can be uncleared. The energy surplus and shortage of LAMs are cleared in the WAM. Thanks to the wide-area two-layer structure, the market outcome is near-social-optimal in large-scale systems. However, the proposed market forms a complex mathematical program with equilibrium constraints (MPEC). To solve the problem, we propose an efficient and hierarchically distributed bidding algorithm. The proposed two-layer market and bidding algorithm are verified on the IEEE 123-bus system with 11250 prosumers, which demonstrates the practicality and efficiency for large-scale markets.
\end{abstract}


\section{Introduction}
\subsection{Background}
In distribution systems, the number of distributed energy resources, e.g., electric vehicles, distributed energy storages, and photovoltaic panels, is rapidly increasing \cite{liu2022merging}. The role of the end-user in the demand side has been transformed from a passive consumer into an active prosumer, which can produce or consume electricity \cite{chen2022review}. Over decades, there have been numerous researches on managing prosumers, for example, energy management \cite{su2023hierarchically}, demand response \cite{su2022multi}, and micro-grids \cite{wang2020asynchronous}. However, most works treat prosumers as controllable resources, ignoring that their interests are not identical to the distribution system operator.

Inspired by the sharing economy principle, energy sharing markets \cite{liu2017energy, li2021operator, fleischhacker2018sharing, wang2014game, morstyn2018bilateral, he2020community, tanaka2018relationship, chen2019energy, le2020peer, wang2021distributed, chen2022energy, guerrero2018decentralized, liu2018hybrid, han2021estimation, mei2019coalitional, han2019improving, el2017managing, zheng2022peer, wang2016incentivizing, wang2018incentive} are proposed to provide prosumers with more incentives. It is inspired by the the sharing economy principle that idle resources are shared to promote social efficiency. In the sharing market, energy producers benefit from selling excessive electricity, while energy consumers can purchase low-priced electricity. Due to the uneven distribution of energy resources, especially renewable energy, it has potential to promote social efficiency by wide-area sharing energy.

\subsection{Related Works}
The initial form of energy sharing markets tends to simplify the prosumer model. Plenty of literature \cite{liu2017energy, li2021operator, fleischhacker2018sharing} regard the prosumer as a price-taker, where the market is commonly modeled as a Stackelberg game. A market operator who sets price acts as a leader, while prosumers make transactive decisions as followers according to the price. Ref. \cite{liu2017energy} proposes a sharing market among photovoltaic prosumers, where the market operator makes different prices for buying and selling. In Ref. \cite{li2021operator}, a day-ahead market among energy storage systems is formulated under the time-of-use price. The leader-follower model may harm the incentives of prosumers and leak their privacies. Numerous researchers turn to separate prosumers into pure sellers and buyers \cite{wang2014game, morstyn2018bilateral, he2020community}, where the double-side auction mechanism is commonly utilized for clearing. In Ref. \cite{morstyn2018bilateral}, a prosumer-generator double auction market with suppliers as intermediaries is proposed for joint real-time and forward markets. Ref. \cite{he2020community} proposes an adaptive bidding strategy to infer the supply and demand from historical trading records under a double auction market of energy storages. In the double-side auction market, the flexibilities of prosumers are suppressed and network constraints are difficult to consider.

In recent years, a broader literature \cite{tanaka2018relationship, chen2019energy, le2020peer} concentrates on unlocking the flexibility in the sharing market. Prosumers are price-makers and they can freely choose to be a buyer or a seller according to the market clearing price. Ref. \cite{tanaka2018relationship} proposes an incentive-compatible market based on the Vickrey-Clarke-Groves mechanism, where the optimal strategies of market participants are willing to report their true cost functions. In Ref. \cite{chen2019energy}, a novel energy sharing mechanism for prosumers is proposed, where the role of the prosumer is endogenously determined after the market clears. Ref. \cite{le2020peer} develops a distributed peer-to-peer market, whose equilibrium is proved equivalent to the social optimum.

Energy sharing under network constraints has attracted much attention in recent years \cite{wang2021distributed, chen2022energy, guerrero2018decentralized}. In theory, \cite{liu2007impacts} proves that the generalized Nash equilibrium of markets under network constraints may be multiple or non-existent. Ref. \cite{wang2021distributed} complements network constraints in the market model in \cite{chen2019energy} and then proposes a distributed generalized Nash equilibrium seeking algorithm. Ref.\cite{chen2022energy} further reveals that some prosumers in the sharing market may strengthen their market power due to network congestion. In Ref. \cite{guerrero2018decentralized}, a continuous double auction market in the low-voltage network is proposed with a sensitivity-based method. In this paper, we design a simple pricing mechanism inspired by the principle of distribution locational marginal pricing (DLMP) \cite{bai2017distribution}, which has potential to be embedded by existing DLMP methods.

Most of the above works focus on small-scale energy sharing markets, whereas prosumers in distribution systems are becoming more and more fragmented and massive. As the number of prosumers participating in the sharing market explodes, the market clearings in the above works may suffer from incredible computational burdens. For instance, in cooperative-game-based markets \cite{liu2018hybrid, han2021estimation, mei2019coalitional}, the number of problems to be solved for calculating Sharpley values exponentially increases with the increasing of prosumers. Ref. \cite{han2019improving} adopts the K-means clustering method to improve the scalability of nucleolus calculation at the cost of inexactness. Stackelberg-game-based market models \cite{liu2017energy, li2021operator, fleischhacker2018sharing} are commonly transformed into mixed-integer programming (MIP) by the big-M method. As $\mathcal{N}\mathcal{P}$-hard problems, large-scale MIPs always suffer from the curse of dimensionality. Therefore, existing works are difficult to implement in the wide-area energy sharing over massive prosumers. In this paper, we will propose a novel two-layer energy sharing market, which can efficiently share the energy of massive prosumers in a wide area.

\subsection{Contributions}
The major contributions of this paper are two-fold:
\begin{enumerate}
\item \textbf{Sharing Scheme}. We propose a novel two-layer energy sharing market, where massive prosumers can share in a wide-area range. In the lower-level local-area market (LAM), prosumers share energy with each other in the same community, where the energy may be not cleared. Then the LAM operators participate in the upper-level wide-area market (WAM) to share uncleared energy. We prove that the market clearing result is unique and approximate to the social optimum in large-scale markets.

\item \textbf{Bidding Algorithm}. The proposed energy sharing market problem belongs to a mathematical program with equilibrium constraints (MPEC). Existing works \cite{liu2017energy, li2021operator, fleischhacker2018sharing, el2017managing, zheng2022peer} commonly transform MPEC into MIP, which is $\mathcal{N}\mathcal{P}$-hard and inefficient in large-scale markets. We first reveal that the equivalence of the Nash equilibrium in the LAM equals to the solution to a convex optimization problem. Then we propose a hierarchically distributed bidding algorithm without binary variables to efficiently clear the whole market.

\end{enumerate}

\subsection{Organization}
The rest of this paper is organized as follows. Section II establishes the two-layer energy sharing framework. The lower-level LAM and upper-level WAM are formulated in Sections III and IV, respectively. Section V verifies the effectiveness and efficiency of the proposed sharing market in a large-scale market. Section VI concludes the paper.

\section{Overall Market Architecture}
Suppose that there are a set of local communities $\mathcal{N}$, indexed by $i \in \mathcal{N}$. In an arbitrary community $i$, a set of prosumers indexed by $j \in \mathcal{U}_i$ will share energy in an LAM $i$. Assume that every prosumer is organized by \textit{one and only one} local community, i.e., $\mathcal{U}_i \cap \mathcal{U}_{i'} = \emptyset, \forall i, i' \in \mathcal{N}, i \neq i'$. There is a WAM for LAM operators to share energy in a wider area.

\begin{figure}[t]
\centering
\includegraphics[width=0.45\textwidth]{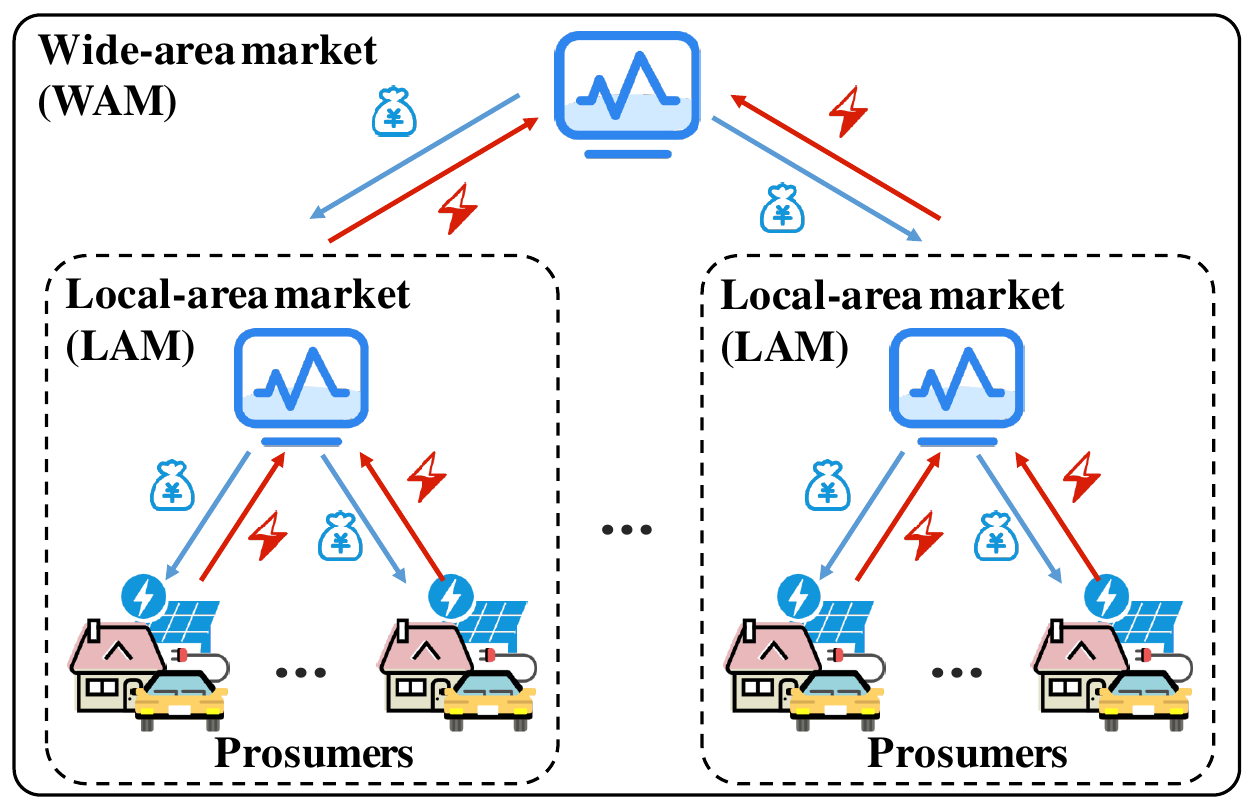}
\caption{Two-layer architecture of the energy sharing market.}
\label{layered_market}
\end{figure}

The two-layer architecture of the proposed sharing market is shown in Fig. \ref{layered_market}. The main idea of the two-layer energy sharing mechanism is as follows. Prosumers in the same LAM first trade with each other. The LAM aims to promote local transactions in each community. Different from existing works\cite{liu2017energy, li2021operator, fleischhacker2018sharing, wang2014game, morstyn2018bilateral, he2020community, tanaka2018relationship, chen2019energy, le2020peer, wang2021distributed, chen2022energy, guerrero2018decentralized, liu2018hybrid, han2021estimation, mei2019coalitional, han2019improving, el2017managing, zheng2022peer, wang2016incentivizing, wang2018incentive}, the LAM does not strictly pursue energy clearing. In our model, a WAM is developed for LAM operators to share unclear energy. The WAM will consider the energy balance and network congestion constraints.

Due to the uneven distribution of renewable energy production and load, the prosumers in the same community may be simultaneously willing to sell or buy electricity. Mandatory clearing in the LAM will damage prosumers' interests. Our model provides a new paradigm, where the uncleared energy can be shared in the upper-level WAM. Social efficiency is promoted since energy is more reasonably distributed.

\section{Lower-Level Local-Area Sharing Market}
\subsection{Energy Sharing Mechanism}
The lower-level local-area sharing market of prosumers in the same local community is managed by the LAM operator. Assume that every prosumer has a distributed generator to generate electricity. Let $p_j$ and $D_j$ denote the power generation and the constant demand of prosumer $j$. The energy sharing market provides a platform for electricity transactions. The power balance equation for each prosumer $j$ is
\begin{align} \label{prosumer_old_model}
D_j + x_j = p_j,
\end{align}
where $x_j$ is the shared energy amount. $x_j > 0$ if prosumer $j$ is a seller, while $x_j < 0$ if it is a buyer.

Most studies \cite{liu2017energy, li2021operator, fleischhacker2018sharing, wang2014game, morstyn2018bilateral, he2020community, tanaka2018relationship, chen2019energy, le2020peer, wang2021distributed, chen2022energy, guerrero2018decentralized, liu2018hybrid, han2021estimation, mei2019coalitional, han2019improving, el2017managing, zheng2022peer} follow similar prosumer models as \eqref{prosumer_old_model}. However, they neglect an important fact that prosumers can purchase energy from electric utilities in the main grid. In fact, it is necessary to consider the effect of the electric utility on prosumers, when studying the energy sharing market. As prosumer $i$ can buy/sell electricity from/to the electric utility, the power balance equation \eqref{prosumer_old_model} is replaced with
\begin{align} \label{prosumer_balance_model}
D_j + x_j + p_j^- = p_j + p_j^+,
\end{align}
where $p_j^+$ and $p_j^-$ are the purchased and sold electricity, respectively.

The strategy of a prosumer is the amount of power generation and the transactions with the electric utility and the energy sharing market. The cost of power generation is commonly quadratic and independent on other prosumers. The buying/selling prices from/to the electric utility, denoted by $\varpi^+$ and $\varpi^-$, are fixed and identical for all prosumers. The energy sharing price in the LAM is decided by all prosumers in the same community, i.e., $\forall j \in \mathcal{U}_i$. Let $\varpi_i ( \bm{x}^{i} )$ denote the energy sharing price in LAM $i$, where $\bm{x}^i$ is the compact form of the collection $\left\{ x_j, \forall j \in \mathcal{U}_i \right\}$.

Therefore, the decision model of prosumer $j$ can be formulated by the following optimization problem
\begin{subequations} \label{prosumer_model}
\begin{align}
\min_{p_j, p_j^+, p_j^-, x_j} ~ & \frac{c_j}{2} p_j^2 + b_j p_j + \varpi^+ p_j^+ - \varpi^- p_j^- - \varpi_i ( \bm{x}^{i} ) x_j \\
\text{s.t.} ~~ & \underline{P}_j \le p_j \le \overline{P}_j, ~~:~ \underline{\mu}_j, \overline{\mu}_j \\
& p_j^+ \ge 0, ~~ p_j^- \ge 0, ~~:~ \mu_j^+, \mu_j^- \\
& D_j + x_j + p_j^- = p_j + p_j^+, ~~:~ \mu_j^{PB}, \label{prosumer_power_balance}
\end{align}
\end{subequations}
where $c_j > 0$ and $b_j$ are constant parameters of the power generation cost, $\underline{P}_j$ and $\overline{P}_j$ are the lower and upper bounds of power generation with $\underline{P}_j \le \overline{P}_j$, and $\underline{\mu}_j, \overline{\mu}_j, \mu_j^+, \mu_j^-$, and $\mu_j^{PB}$ are Lagrangian multipliers with respect to these constraints.

The energy sharing price $\varpi_i ( \bm{x}^i )$ is defined as \cite{el2017managing}
\begin{align} \label{CESM_pricing}
\varpi_i ( \bm{x}^i ) = \varpi_i^0 - a_i \sum_{j\in\mathcal{U}_i} x_j,
\end{align}
where $\varpi_i^0$ is the base price and $a_i > 0$ is the price elasticity. $\sum_{j\in\mathcal{U}_i} x_j$ indicates the supply-demand relationship. The base price $\varpi_i^0$ is decided by the community market operator, while the real energy sharing price depends on the supply-demand relationship. The oversupply of energy in the LAM, i.e., $\sum_{j\in\mathcal{U}_i} x_j > 0$, leads to price reduction; otherwise, the sharing price rises.

Different from the LAM, the buying/selling price from/to the electric utility is fixed. Electric utilities are more inclined to sell electricity rather than buy it since bought energy fluctuates and increases the operation burden. Selling electricity is a mandatory responsibility for electric utilities, while buying from prosumers is not necessary. Besides, the electric utility always plays the role of a price-maker due to the supply-demand relationship. Thus electric utilities intend to arbitrage from electricity sellers and buyers by setting different buying and selling prices. It is widely adopted in electricity markets \cite{wang2016incentivizing, wang2018incentive}. In economics, this phenomenon is called price discrimination. 

\begin{assumption}[Price discrimination] \label{assum_price_discrimination}
The prices of the electricity utility satisfies $\varpi^+ > \varpi^- > 0$.
\end{assumption}

In the proposed sharing market, price discrimination prevents prosumers from arbitraging from the electric utility. A prosumer will either buy electricity from the electric utility or sell to it, but not both. See the following proposition.

\begin{proposition} \label{prop_complementary}
Under Assumption \ref{assum_price_discrimination}, given $x_{j'}, \forall j'\in \mathcal{U}_i \setminus \left\{j\right\}$, the optimal purchased and sold electricity are complementary, i.e., $p_j^+ p_j^- = 0$.
\end{proposition}

The proof of Proposition \ref{prop_complementary} can be found in Appendix \ref{appendix_complementary}. Then we prove that prosumers are willing to participate in the LAM to share energy. Let $\mathcal{F}_j^{-x}$ denote the optimal cost of prosumer $j$ that only trades with the electric utility.

Then we will show the incentives of prosumers to participate in the LAM even when the electric utility can trade with them.

\begin{proposition} \label{prop_incentive}
Under Assumption \ref{assum_price_discrimination}, we have for $\forall j \in \mathcal{U}_i$
\begin{align}
\mathcal{F}_j (s_j^*, \bm{s}_{-j}^*) \le \mathcal{F}_j^{-x}.
\end{align}
\end{proposition}
Proposition \ref{prop_incentive} holds directly since $\mathcal{F}_j^{-x}$ is the optimal value of problem \eqref{prosumer_model} with an extra constraint $x_j = 0$.

\subsection{Energy Sharing as A Nash Game}
The prosumer model \eqref{prosumer_model} constitutes a Nash game, where prosumers' payoffs depend on others in the same community. Denote by $s_j$ the collection of the decision variables $p_j, p_j^+, p_j^-, x_j$ of prosumer $j$. Let $\bm{s}_{-j}$ denote the compact form of prosumers in community set $\mathcal{U}_i$ except $j$, i.e., $\mathcal{U}_i \setminus \left\{j\right\}$. The payoff and action set of prosumer $j$ can be denoted by $\mathcal{F}_j (s_j, \bm{s}_{-j})$ and $\mathcal{S}_j$, respectively. The Nash game model \eqref{prosumer_model} can be simplified by
\begin{align} \label{Nash_game}
\min_{s_j \in \mathcal{S}_j} ~ \mathcal{F}_j (s_j, \bm{s}_{-j}).
\end{align}

The Nash game is composed of the following three elements:
\begin{enumerate}
\item the set of players $\mathcal{U}_i$;

\item action sets $\mathcal{S}_j, \forall j\in \mathcal{U}_i$, and the strategy space $\mathcal{S}^i := \prod_{j\in\mathcal{U}_i} \mathcal{S}_j$;

\item payoffs $\mathcal{F}_j (s_j, \bm{s}_{-j}), \forall j\in \mathcal{U}_i$.
\end{enumerate}

Then we will characterize some properties of the equilibrium of the LAM.

\begin{definition}[Nash Equilibrium] \label{defi_Nash_equilibrium}
A strategy profile $\widehat{\bm{s}}^i := \left\{ \widehat{s}_j, \forall j\in\mathcal{U}_i \right\}$ is a Nash equilibrium (NE) of the Nash game \eqref{Nash_game}, if
\begin{align}
\mathcal{F}_j (\widehat{s}_j, \widehat{\bm{s}}_{-j}) \le \mathcal{F}_j (s_j, \widehat{\bm{s}}_{-j}),~ \forall s_j \in \mathcal{S}_j, j \in \mathcal{U}_i.
\end{align}
\end{definition}

In this paper, let $\bm{s}^{i*} = \left\{ p_j^*, p_j^{+*}, p_j^{-*}, x_j^*, \forall j \in \mathcal{U}_i \right\}$ denote an NE of the Nash game \eqref{Nash_game} in LAM $i$. The following proposition proves the uniqueness of the NE.

\begin{proposition} \label{prop_NE_equivalent_solution}
Under Assumption \ref{assum_price_discrimination}, the Nash game \eqref{Nash_game} exists a unique NE $\bm{s}^{i*}$, which is also the unique optimal solution to the following problem.
\begin{subequations} \label{equivalent_problem}
\begin{align}
\min_{p_j, p_j^+, p_j^-, x_j} & \sum_{j\in\mathcal{U}_i} \Big\{ \frac{c_j}{2} p_j^2 + b_j p_j + \varpi^+ p_j^+ - \varpi^- p_j^- - \varpi_i^0 x_j \Big\}  \notag \\
& + \frac{a_i}{2} \Big(\sum_{j\in\mathcal{U}_i} x_j \Big)^2 + \frac{a_i}{2} \sum_{j\in\mathcal{U}_i} x_j^2 \\
\text{s.t.} ~~ & \underline{P}_j \le p_j \le \overline{P}_j ~~:~ \underline{\mu}_j, \overline{\mu}_j, ~\forall j \in \mathcal{U}_i \label{cons_power_prod} \\
& p_j^+ \ge 0, ~~ p_j^- \ge 0 ~~:~ \mu_j^+, \mu_j^-, ~\forall j \in \mathcal{U}_i \\
& D_j + x_j + p_j^- = p_j + p_j^+ ~:~ \mu_j^{PB}, \forall j \in \mathcal{U}_i. \label{power_balance_equivalent}
\end{align}
\end{subequations}
\end{proposition}

The proof of Proposition \ref{prop_NE_equivalent_solution} can be found in Appendix \ref{appendix_NE_equivalent_solution}. Proposition \ref{prop_NE_equivalent_solution} indicates that the Nash game of the LAM is well defined.

\subsection{Bidding Algorithm}
The bidding algorithm of the LAM is presented in Algorithm \ref{algo_CESM_bidding}. Prosumers only need to submit the decision of shared energy to the market. Their private information, e.g., the energy demand or the generation parameters, will not be leaked to the market operator or other prosumers. Prosumer's problem \eqref{prosumer_prob_iteration} is equivalent to problem \eqref{prosumer_model} with $x_{j'} \equiv x_{j'} (h-1), \forall j'\in \mathcal{U}_i \setminus \left\{j\right\}$. Note that each prosumer does not have to communicate with others. It can directly deduce $\sum_{j'\in \mathcal{U}_i \setminus \left\{j\right\}} x_{j'}$ by $\varpi_i (h-1)$. After solving \eqref{prosumer_prob_iteration}, the real bidding is averaged by the last one as \eqref{prosumer_prob_iteration2}, which describes prosumers' inertia in the decision. From the perspective of the algorithm, the step size $\rho_i$ decides the convergence of the bidding.

\begin{algorithm}[t]
\caption{Bidding Algorithm of LAM $i$} 
\label{algo_CESM_bidding}

\hangafter 1
\hangindent 1em
\textbf{Input:} Error tolerance $\varepsilon_i$, base price $\varpi_i^0$, price elasticity $a_i$, initial energy sharing price $\varpi_i (0)$, initial shared energy $x_j (0), \forall j \in \mathcal{U}_i$, step size $\rho_i \in (0, 1)$, and iteration index $h = 1$.

\hangafter 1
\hangindent 1em
\textbf{Output:} Prosumer strategy $p_j^*, p_j^{+*}, p_j^{-*}, x_j^*, \forall j\in\mathcal{U}_i$ and energy sharing price $\varpi_i ( \bm{x}^{i*} )$.

\hangafter 1
\hangindent 1em
\textbf{S1:} Each prosumer $j\in\mathcal{U}_i$ decides a sharing strategy $s_j (h) = \left\{ p_j (h), p_j^+ (h), p_j^- (h), x_j (h) \right\}$ with $\varpi_i (h-1)$ by solving the following problem
\begin{subequations}
\begin{align}
\widetilde{s}_j (h) =& \arg \min_{s_j \in \mathcal{S}_j} ~  \frac{c_j}{2} p_j^2 + b_j p_j + \varpi^+ p_j^+ - \varpi^- p_j^- \notag \\
& - [\varpi_i (h-1) + a_i x_j(h-1) - a_i x_j] x_j \label{prosumer_prob_iteration} \\
s_j (h) =& ~\rho_i \widetilde{s}_j (h) + (1 - \rho_i) s_j (h-1) \label{prosumer_prob_iteration2}
\end{align}
\end{subequations}
and then send the shared energy $x_j (h)$ to the LAM. 

\hangafter 1
\hangindent 1em
\textbf{S2:} After receiving all $x_j (h), j\in\mathcal{U}_i$, the LAM operator readjusts the energy sharing price by
\begin{align}
\varpi_i (h) &= \varpi_i^0 - a_i \sum_{j\in\mathcal{U}_i} x_j (h)
\end{align}
and then broadcasts it to all prosumers.

\hangafter 1
\hangindent 1em
\textbf{S3:} If $\left| \varpi_i (h) - \varpi_i (h-1) \right| \le \varepsilon_i$, terminate the bidding and output the latest strategy and price; otherwise, set $h \leftarrow h + 1$ and go to \textbf{S1}.
\end{algorithm}

\subsection{Prices in Equilibrium}
Existing works commonly assume that prosumers can only trade in the energy market. However, they ignore the influence of electric utilities on sharing markets. We will analyze the condition where prosumers can share energy with each other in the sharing market or/and trade with the electric utility. The interaction among the sharing price $\varpi_i ( \bm{x}^i )$, the buying/selling prices $\varpi^+, \varpi^-$, and the shadow prices of prosumers are revealed. To this end, we present the definition of shadow price quantifying the marginal consumption/production price of the prosumer.

\begin{definition}[Shadow Price] \label{defi_shadow_price}
The shadow price of prosumer $j$, $\mu_j^{PB*}$, is defined as the optimal Lagrangian multiplier $\mu_j^{PB}$ with respect to the power balance constraint \eqref{prosumer_power_balance} at the NE.
\end{definition}

The relationship among the shadow price, the electric utility price, and the energy sharing price is clarified below.

\begin{proposition} \label{prop_pricing_relationship}
Under Assumption \ref{assum_price_discrimination}, at the NE $\bm{s}^{i*} = \left\{ p_j^*, p_j^{+*}, p_j^{-*}, x_j^*, \forall j \in \mathcal{U}_i \right\}$, we have
\begin{enumerate}[label=\alph*)]
\item the shadow price of each prosumer $j \in \mathcal{U}_i$ satisfies
\begin{align}
\varpi^- \le \mu_j^{PB*} \le \varpi^+.
\end{align}

\item the shared energy is given by
\begin{align} \label{prosumer_role}
x_j^* = \frac{\varpi_i ( \bm{x}^{i*} ) - \mu_j^{PB*}}{a_i}.
\end{align}

\item the energy sharing price at equilibrium satisfies
\begin{align} \label{sharing_price_equation}
\varpi_i ( \bm{x}^{i*} ) = \frac{\varpi_i^0 + \sum_{j\in \mathcal{U}_i} \mu_j^{PB*}}{1 + |\mathcal{U}_i|}.
\end{align}

\item if the base price $\varpi_i^0$ in the LAM \eqref{CESM_pricing} satisfies
\begin{align} \label{moderate_base_price}
\varpi^- \le \varpi_i^0 \le \varpi^+,
\end{align}
then
\begin{align} \label{sharing_price_bound}
\varpi^- \le \varpi_i ( \bm{x}^{i*} ) \le \varpi^+. 
\end{align}
Especially, even if the condition \eqref{moderate_base_price} is not satisfied, sufficiently numerous prosumers also yield \eqref{sharing_price_bound}.
\end{enumerate}
\end{proposition}

The proof of Proposition \ref{prop_pricing_relationship} can be found in Appendix \ref{appendix_pricing_relationship}. Proposition \ref{prop_pricing_relationship} reveals the competitiveness of the LAM against the electric utility. The energy sharing price $\varpi_i ( \bm{x}^{i*} )$ are commonly located in the range $\left[\varpi^-, \varpi^+\right]$. It means that prosumers can buy electricity at a lower price than $\varpi^+$ and sell at a higher price than $\varpi^-$. That is the real reason why prosumers are willing to participate in the LAM. Besides, from the perspective of the market, the interest of the electric utility by price discrimination is cut down due to the competition of the LAM. Because of the existence of the sharing market, the electric utility cannot severely set discriminative prices.

\begin{figure}[t]
\centering
\includegraphics[width=0.45\textwidth]{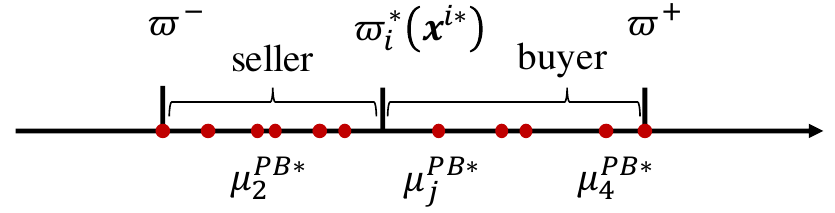}
\caption{An illustration of the relationship among prices.}
\label{price_relationship}
\end{figure}

Equation \eqref{prosumer_role} indicates that the role of a prosumer depends on the values of the energy sharing price and the shadow price. If the energy sharing price $\varpi_i ( \bm{x}^{i*} )$ is larger than the shadow price $\mu_j^{PB*}$, the prosumer will sell energy, i.e., $x_j^* > 0$. Otherwise, the prosumer will be an energy buyer.

Fig. \ref{price_relationship} illustrates the relationship among prices. The shadow price of a prosumer guarantees its role in the transaction. Prosumers with a lower shadow price tend to sell electricity to those with a higher shadow price. The real energy sharing price $\varpi_i ( \bm{x}^{i*} )$ is the average of these sharing prices with the base price.

\section{Upper-Level Wide-Area Sharing Market}
\subsection{Energy Sharing Mechanism}
Recall that in each LAM, the shared energy does not have to be cleared. These community market operators will share the uncleared energy in the upper-level WAM managed by the WAM operator. In the WAM, an LAM operator can be regarded as an aggregated ``prosumer", whose uncleared energy is the shared electricity in the WAM.

In the WAM, the energy shared by LAMs is traded at the base price $\varpi_i^0$. The WAM operator can adjust the base price to influence the equilibrium of LAMs. Let $x_j^* \left( \varpi_i^0 \right)$ denote the shared energy of prosumer $j$ at the NE under the base price $\varpi_i^0$. Define the uncleared energy of community $i$, i.e., the shared energy in the WAM, as
\begin{align} \label{y_defi}
y_i \left( \varpi_i^0 \right) = \sum_{j\in \mathcal{U}_i} x_j^* \left( \varpi_i^0 \right).
\end{align}
Equation \eqref{y_defi} aggregates the volume-price curve of an LAM.

The clearing rule of the WAM is solving the following problem to decide base prices in all LAMs
\begin{subequations} \label{DESM_model}
\begin{align}
\min_{\bm{\varpi}^0} ~ & f(\bm{\varpi}^0) \\
\text{s.t.} ~~ & \bm{\varpi}^0 := \left\{ \varpi_i^0, i \in \mathcal{N} \right\} \notag \\
& \sum_{i\in\mathcal{N}} y_i \left( \varpi_i^0 \right) = 0 \label{DESM_power_balance} \\
& \sum_{i\in\mathcal{N}} \pi_{il} ~ y_i \left( \varpi_i^0 \right) \le F_l, ~\forall l \in \mathcal{L}, \label{DESM_network_cons}
\end{align}
\end{subequations}
where $\mathcal{L}$ is the set of network constraints, $\pi_{il}$ is the parameter quantifying the factor of LAM $i$ to network constraint $l$, $F_l$ is the constant physical limitation of network constraint $l$. Constraint \eqref{DESM_power_balance} is the energy clearing limitation, while constraint \eqref{DESM_network_cons} guarantees the power flow security under network constraints.

The objective $f(\bm{\varpi}^0)$ can be some function of social welfare. However, purely attaining a feasible solution to the problem is extremely hard. Noting that every $y_i \left( \varpi_i^0 \right)$ implies an NE of an LAM, problem \eqref{DESM_model} belongs to an MPEC. To solve it, integer variables are commonly introduced to simplify complementary slackness conditions in equilibriums, which transforms the MPEC to a mixed-integer program (MIP) \cite{liu2017energy, li2021operator, fleischhacker2018sharing, el2017managing, zheng2022peer}. There are massive prosumers in the energy sharing problem. An enormous number of integer variables are required for transforming complementary slackness conditions \eqref{KKT_mu_p1}-\eqref{KKT_mu_p_minus}, which leads to the curse of dimensionality in solving the MIP problem.

To this end, we propose a novel bidding algorithm for the WAM clearing problem. Then we prove that the algorithm can attain a feasible price strategy, which simultaneously approaches to the social optimal.

\subsection{Bidding Algorithm}
The basic idea is to linearly adjust the sharing price according to the violation level of the power balance and network congestion constraints. To this end, we first analyze the property of $y_i \left( \varpi_i^0 \right)$, yielding the following proposition.

\begin{proposition} \label{prop_monotonicity}
Under Assumption \ref{assum_price_discrimination}, the uncleared energy $y_i \left( \varpi_i^0 \right)$ in each LAM $i \in \mathcal{N}$ is monotonically non-decreasing with respect to $\varpi_i^0$.
\end{proposition}
The proof of Proposition \ref{prop_monotonicity} can be found in Appendix \ref{appendix_monotonicity}. Based on this, we define the base price as
\begin{align}
\varpi_i^0 = \varpi^{PB} + \varpi_i^{NET} = \varpi^{PB} + \sum_{l\in\mathcal{L}} \pi_{il} \varpi_l,
\end{align}
where $\varpi^{PB}$ is the basic sharing price based on the power balance, $\varpi_i^{NET}$ is the additional price on LAM $i$ according to the network congestions, $\varpi_l$ is the congestion price of network constraint $l$, and parameter $\pi_{il}$ quantifies the sensitivity of energy sharing of LAM $i$ on network constraint $l$.

According to the monotonicity of $y_i \left( \varpi_i^0 \right)$, the prices are updated as
\begin{subequations} \label{DESM_bidding}
\begin{align}
\varpi^{PB} (k+1) &= \varpi^{PB} (k) - \alpha^{PB} \sum_{i\in\mathcal{N}} y_i \left( \varpi_i^0 (k) \right) \\
\varpi_l (k+1) &= \left[ \varpi_{l} (k) - \alpha_l \left( \sum_{i\in\mathcal{N}} \pi_{il} ~ y_i \left( \varpi_i^0 (k) \right) - F_l \right) \right]^-
\end{align}
\end{subequations}
where $\left[\cdot\right]^-$ is the projection onto the range $\left( -\infty, 0\right]$, $\alpha^{PB}, \alpha_l > 0, \forall l\in\mathcal{L}$ are step sizes. The complete bidding algorithm is presented in Algorithm \ref{algo_DESM_bidding}.

\begin{algorithm}[t]
\caption{Bidding Algorithm of the WAM} 
\label{algo_DESM_bidding}

\hangafter 1
\hangindent 1em
\textbf{Input:} Error tolerance $\varepsilon$, initial energy sharing price $\varpi^{PB} (0)$, initial network congestion prices $\varpi_l (0), \forall l\in\mathcal{L}$, initial base price $\varpi_i^0 (0) = \varpi^{PB} (0) + \sum_{l\in\mathcal{L}} \pi_{il} \varpi_l (0), \forall i\in\mathcal{N}$, step sizes $\alpha^{PB}, \alpha_l > 0$, and iteration index $k = 0$.

\hangafter 1
\hangindent 1em
\textbf{Output:} Shared energy $y_i^*, \forall i\in\mathcal{N}$ and clearing prices $\varpi^{PB*}, \varpi_i^{NET*}, \varpi_i^{0*}, \forall i\in\mathcal{N}$.

\hangafter 1
\hangindent 1em
\textbf{S1:} Given the base price $\varpi_i^0 (k)$, each LAM $i\in\mathcal{N}$ clears by \textbf{Algorithm \ref{algo_CESM_bidding}} and then sends the shared energy $y_i \left( \varpi_i^0 (k) \right)$ to the WAM.

\hangafter 1
\hangindent 1em
\textbf{S2:} After receiving all $y_i \left( \varpi_i^0 (k) \right), i\in\mathcal{N}$, the WAM readjusts the energy sharing price by \eqref{DESM_bidding} and then broadcasts prices $\varpi_i^0 (k+1)$ to all LAMs.

\hangafter 1
\hangindent 1em
\textbf{S3:} If $\left| \varpi_i (k) - \varpi_i (k-1) \right| \le \varepsilon, \forall i\in\mathcal{N}$, terminate the bidding and output the latest strategy and prices; otherwise, set $k \leftarrow k + 1$ and go to \textbf{S1}.
\end{algorithm}

\subsection{Social Efficiency}
Then we will prove the convergence of Algorithm \ref{algo_DESM_bidding} and reveal its market efficiency.

\begin{proposition} \label{prop_MPEC}
Under Assumption \ref{assum_price_discrimination}, Algorithm \ref{algo_DESM_bidding} converges to a unique equilibrium $\left\{ p_j^*, p_j^{+*}, p_j^{-*}, x_j^*, y_i^* \forall j \in \mathcal{U}_i, i \in \mathcal{N} \right\}$, which is also the unique optimal solution to the following problem.
\begin{subequations} \label{equivalent_problem_total}
\begin{align}
\min_{p_j, p_j^+, p_j^-, x_j, y_i} & \sum_{i\in\mathcal{N}} \sum_{j\in\mathcal{U}_i} \Big\{ \frac{c_j}{2} p_j^2 + b_j p_j + \varpi^+ p_j^+ - \varpi^- p_j^- \Big\}  \notag \\
& + \sum_{i\in\mathcal{N}} \Big\{ \frac{a_i}{2} y_i^2 + \frac{a_i}{2} \sum_{j\in\mathcal{U}_i} x_j^2 \Big\} \label{equivalent_problem_obj} \\
\text{s.t.} ~~ & \left\{ p_j , p_j^+, p_j^-, x_j \right\} \in \mathcal{S}_j, \forall j\in\mathcal{U}_i, i\in\mathcal{N} \label{cons_prosumer} \\
& y_i = \sum_{j\in \mathcal{U}_i} x_j, \forall i\in\mathcal{N} \label{cons_community} \\
& \sum_{i\in\mathcal{N}} y_i = 0, ~~:~ \lambda^{PB} \label{cons_power_balance} \\
& \sum_{i\in\mathcal{N}} \pi_{il} ~ y_i \le F_l, ~~:~ \lambda_l, ~\forall l \in \mathcal{L} \label{cons_power_flow} 
\end{align}
\end{subequations}
\end{proposition}

The proof of Proposition \ref{prop_MPEC} can be found in Appendix \ref{appendix_MPEC}. Proposition \ref{prop_MPEC} directly shows that the market clear result by Algorithm \ref{algo_DESM_bidding} is a feasible solution to problem \eqref{DESM_model}. As a result, the base price satisfies $\varpi_i^0 = -\lambda^{PB} - \sum_{l\in\mathcal{L}} \pi_{il} \lambda_l$, which can be regarded as the DLMP \cite{bai2017distribution} and may integrate existing DLMP approaches.

Then we will show that the market clear result also performs excellent from the perspective of social welfare. Define the social optimum of the whole system as follows.

\begin{definition}[Social Optimum] \label{defi_social_welfare}
A strategy of prosumers and community market operators $\{p_j, p_j^+, p_j^-, x_j, y_i, j\in\mathcal{U}_i, i\in\mathcal{N}\}$ is socially optimal if it is the solution to the following problem
\begin{subequations} \label{problem_social_welfare}
\begin{align}
\min_{p_j, p_j^+, p_j^-, x_j, y_i} & \sum_{i\in\mathcal{N}} \sum_{j\in\mathcal{U}_i} \Big\{ \frac{c_j}{2} p_j^2 + b_j p_j + \varpi^+ p_j^+ - \varpi^- p_j^- \Big\} \\
\text{s.t.} ~~ & \eqref{cons_prosumer}-\eqref{cons_power_flow}.
\end{align}
\end{subequations}
\end{definition}

\begin{remark}{\textit{(Social Efficiency)}} \label{remark_a}
Note that the only difference between problems \eqref{equivalent_problem_total} and \eqref{problem_social_welfare} is the quadratic terms in the objective function \eqref{equivalent_problem_obj}. The terms can be regarded as the social welfare loss from market competition, whose degree depends on the price elasticities $a_i, \forall i\in\mathcal{N}$. In the case study in Section \ref{social_welfare}, we show that the price elasticities are supposed to decrease with the increasing scale of prosumers. Therefore, it reveals that the market clearing result approximates to the social optimum in large-scale markets.
\end{remark}

\section{Case Study}
\subsection{Setup}
\begin{figure}[t]
\centering
\includegraphics[width=0.45\textwidth]{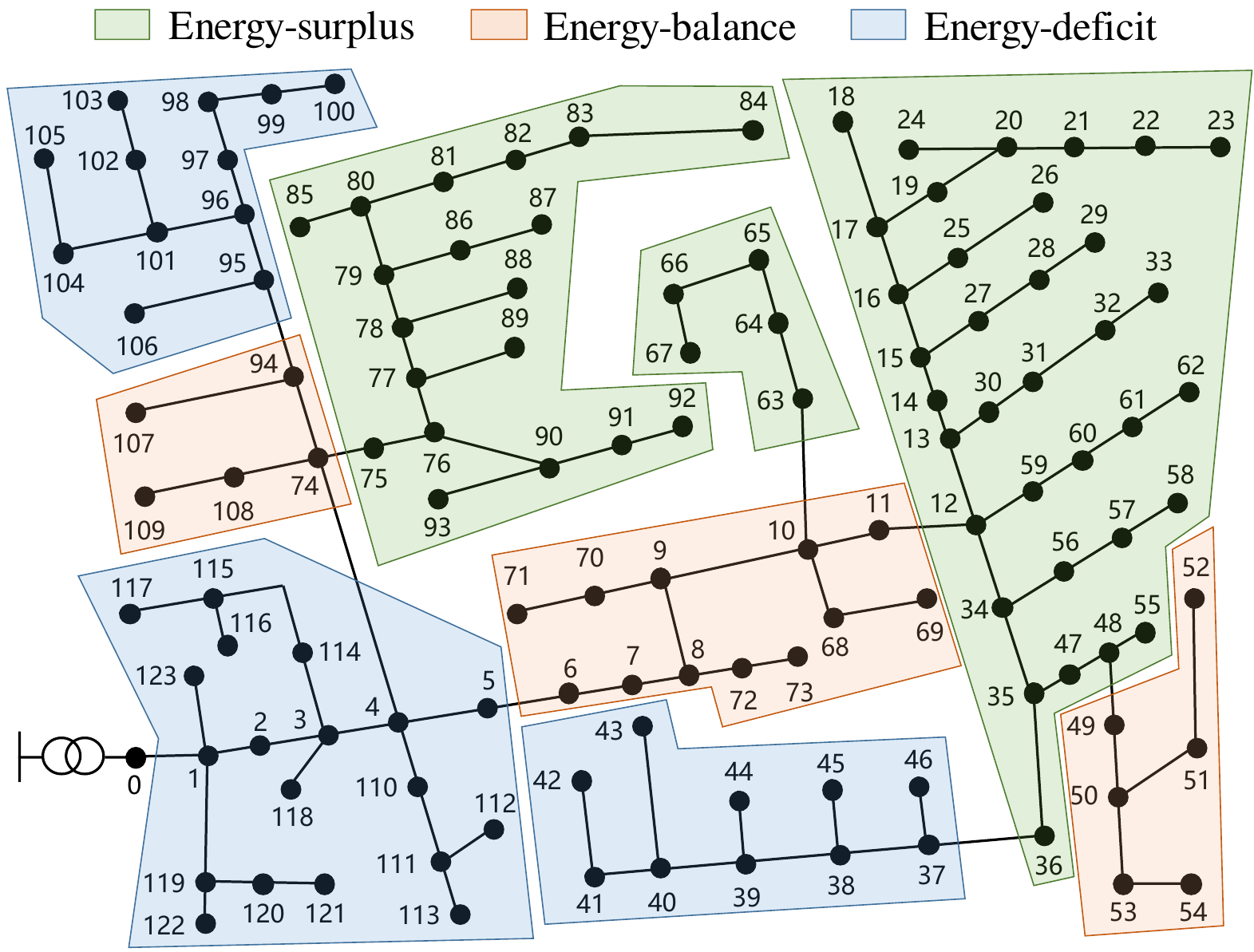}
\caption{Configuration of the IEEE 123-bus system.}
\label{fig_case123}
\end{figure}

In this section, simulation experiments are carried out in MATLAB 2021A on a desktop with an Intel i7-10700 CPU and 16 GB memory. As shown in Fig. \ref{fig_case123}, the IEEE 123-bus system is adopted, where every bus is integrated by an LAM. The number of prosumers in an LAM is in the range of $\left[50, 525\right]$, depending on the raw data of active power at the bus. The total number of LAMs and prosumers are up to 123 and 11250, respectively. The capacities of distribution lines $\left\{10, 63\right\}$, $\left\{35, 36\right\}$, $\left\{4, 74\right\}$, $\left\{8, 9\right\}$, $\left\{15, 16\right\}$, $\left\{78, 79\right\}$, and $\left\{95, 96\right\}$ are 0.5, 12, 1.5, 21, 9, 10, and 15 MW, respectively.

The buying and selling prices $\varpi^+, \varpi^-$ of the electric utility are set to $0.2\$/kW$ and $0.05\$/kW$, respectively. For each LAM, the price elasticity $a_i$ is randomly chosen within the range $\left[2.5, 5\right] \times 10^{-3} / \left|\mathcal{U}_i\right| \$/kW$. The parameters of prosumers are randomly chosen within the ranges: $c_j \in \left[0.5, 1\right] \times 10^{-3} \$/kW^{-2}$, $b_j \in \left[0.01, 0.05\right] \$/kW$, $D_j \in \left[0, 40\right] kW$, and $\underline{P}_j = 0 kW$. Since renewable energy is fluctuating, we design five types of maximal power generations randomly chosen within the ranges $\left[35, 50\right] kW$, $\left[20, 35\right] kW$, $\left[15, 25\right] kW$, $\left[5, 10\right] kW$, and $\left[0, 5\right] kW$, respectively. Then the LAMs are divided into three types: energy- surplus/balance/deficit communities. In energy-surplus LAMs, 80\% and 20\% prosumers pick the maximal power generations from the first and second ranges, respectively. The maximal power generations of prosumers in energy-balance LAMs uniformly fall into the five ranges. In energy-deficit ones, 80\% and 20\% prosumers get their maximal power generations from the fifth and fourth ranges, respectively. Set the step sizes $\alpha^{PB} = 1\times 10^{-6}, \alpha_l = 5\times 10^{-7}, \forall l\in\mathcal{L}$ in \eqref{DESM_bidding}. For Algorithm \ref{algo_CESM_bidding}, we design an adaptive step size with the initial value $\rho_i = 0.2, \forall i\in\mathcal{N}$. During iterations, halve $\rho_i$ if $\varpi_i(h) - \varpi_i(h-1) > 10^{-3} \land \varpi_i(h) - \varpi_i(h+1) > 10^{-3}$ or $\varpi_i(h-1) - \varpi_i(h) > 10^{-3} \land \varpi_i(h+1) - \varpi_i(h) > 10^{-3}$.

\subsection{Main Results}
\begin{figure}[t]
\centering
\includegraphics[width=0.45\textwidth]{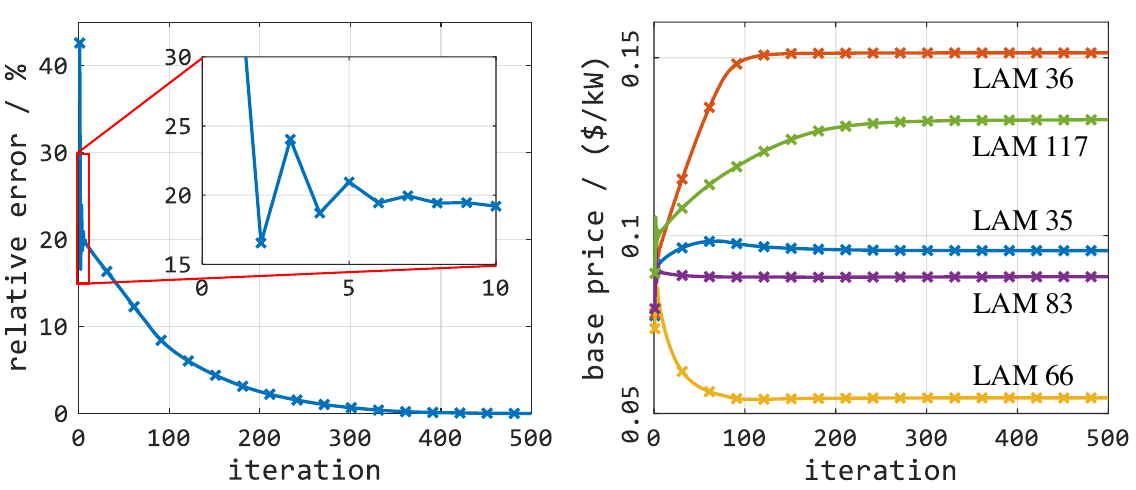}
\caption{Total cost relative error and sharing price during iterations in the WAM.}
\label{fig_Iteration_DESM}
\end{figure}

The WAM clearing process by Algorithm \ref{algo_DESM_bidding} is shown in Fig. \ref{fig_Iteration_DESM}. We calculate the sum of all 11250 prosumers' costs during 500 iterations and record its relative error from the exact value by solving problem \eqref{prop_MPEC} in the left subfigure. It verifies the convergence of Algorithm \ref{algo_DESM_bidding} and the correctness of Proposition \ref{prop_MPEC}. The computational time of Algorithm \ref{algo_DESM_bidding} and solving problem \eqref{prop_MPEC} are 10.2s and 23.8s, respectively. Since the biddings of all prosumers in Algorithm \ref{algo_DESM_bidding} are serially calculated, the average time spent by one prosumer is only $9.07\times 10^{-4}$s, $1/26193$ times of solving problem \eqref{prop_MPEC}. It validates the scalability of the proposed market clearing algorithm in large-scale markets. We pick up five LAMs (35, 36, 66, 83, and 117) to display their base prices $\varpi_i^0$ during iterations in the right subfigure. Due to the congestion of distribution line $\left\{35, 36\right\}$, the base price of LAM 36 is 0.056 \$/kW larger than that of LAM 35.

\begin{figure}[t]
\centering
\includegraphics[width=0.45\textwidth]{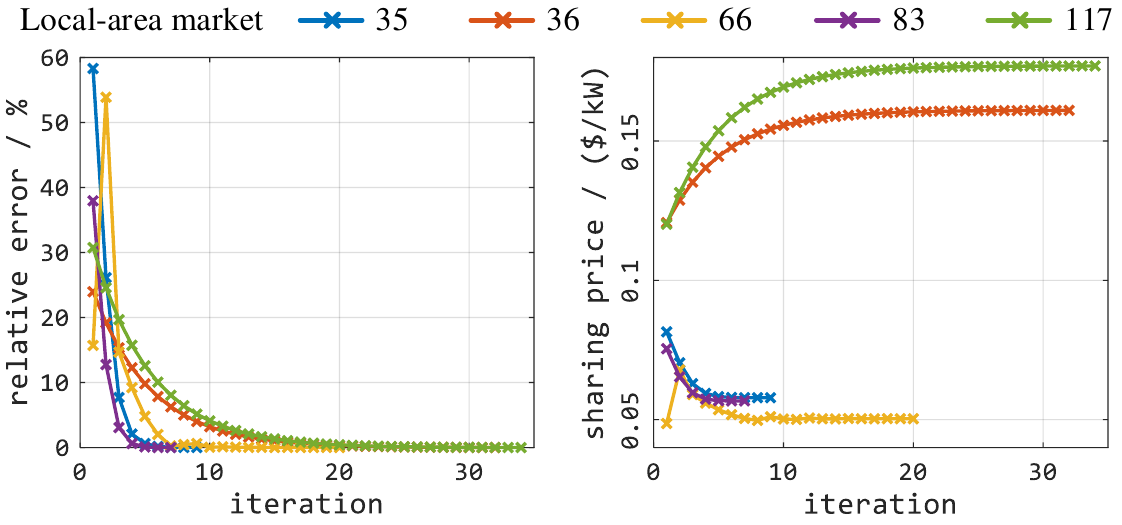}
\caption{Total cost relative errors and sharing prices during iterations in the LAMs.}
\label{fig_Iteration_CESM}
\end{figure}

The market clearing process in the LAMs 35, 36, 66, 83, and 117 by Algorithm \ref{algo_CESM_bidding} during the 25th iteration of the WAM bidding are displayed in Fig. \ref{fig_Iteration_CESM}. In the left subfigure, the total costs of prosumers in one community are calculated and compared to the value obtained by solving problem \eqref{equivalent_problem}, which verifies the convergence of Algorithm \ref{algo_CESM_bidding} and the correctness of Proposition \ref{prop_NE_equivalent_solution}. The right subfigure shows the sharing prices during iterations. The required iteration numbers of LAMs to reach the error tolerance $10^{-8}$ are different. The average number of iterations over the whole algorithm is just 15.1, which is surpassingly quick.

\begin{figure}[t]
\centering
\includegraphics[width=0.45\textwidth]{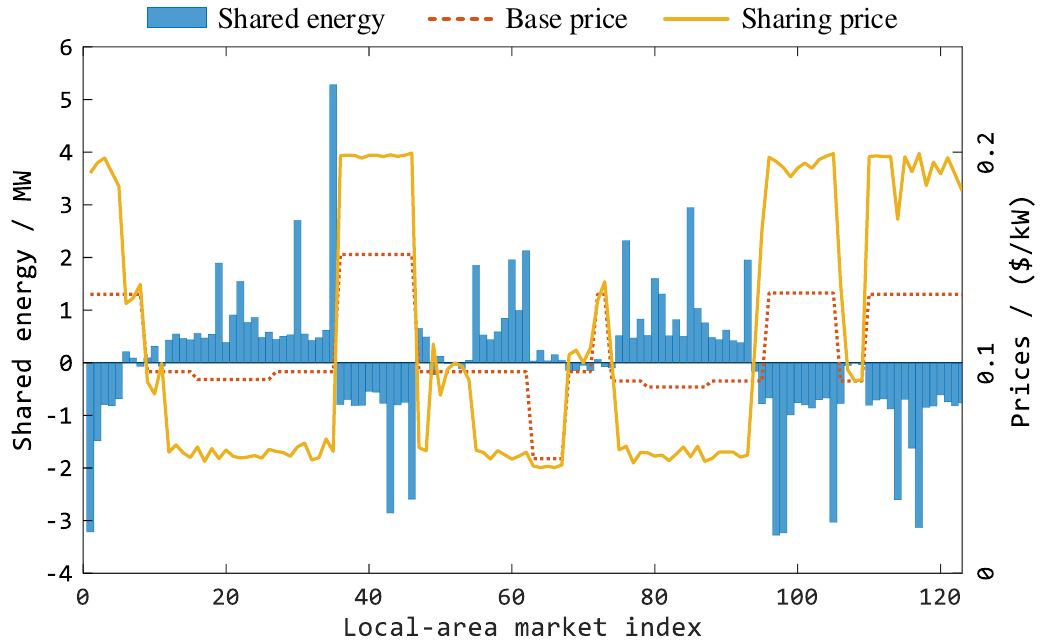}
\caption{Upper-level wide-area sharing market outcome.}
\label{fig_market_clear}
\end{figure}

Fig. \ref{fig_market_clear} shows the WAM outcome. The LAMs vary in the base price due to distribution line congestion. For instance, the base prices of LAMs 63-67 are near to the selling price $\varpi^- = 0.05$ \$/kW since the shared energy reaches the capacity of distribution line $\left\{10, 63\right\}$. Similarly, the base price of LAMs 36-46 are extremely high because of the congestion of line $\left\{35, 36\right\}$. This pricing mechanism can indirectly guide the energy sharing by influencing the base prices. The figure also shows that the final sharing prices are all located within the range of $\left[\varpi^-, \varpi^+\right]$, verifying Proposition \ref{price_relationship}. It reveals the competitiveness of the proposed sharing market against the electric utility: higher selling price and lower purchasing price.

\subsection{Social Efficiency} \label{social_welfare}
We consider four conditions for energy sharing: 1) \textbf{Self-sufficiency (SS)}, energy balance by itself without energy sharing; 2) \textbf{Local-area sharing (LS)}, sharing and clearing energy with prosumers in the same community; 3) \textbf{Local-area optimization (LO)}, local-area cooperation for social optimum; 4) \textbf{Wide-area sharing (WS)}, the proposed two-layer market; and 5) \textbf{Wide-area optimization (WO)}, wide-area cooperation for social optimum. Table \ref{Result} lists the total costs of all prosumers in the four conditions. From SS to LS to WS, with the expansion of the sharing domain, the total cost remarkably reduces, highlighting the significance of wide-area sharing. The local-area sharing LS almost achieves the same total cost as that of local-area optimization LO, i.e., the theoretical lower bound. However, due to the uneven distribution of energy resources, there is still great potential for wide-area sharing from the local- to the wide-area optimization. Therefore, the cost of the proposed two-layer market, i.e., WS, is much lower than that of local-area optimization LO.


\setlength{\tabcolsep}{3mm}{
\begin{table}[t]
\centering
\caption{Total Cost in Different Conditions}
\label{Result}

\begin{tabular}{c c c c c c}
\hline
\textbf{Conditions}	&	\textbf{SS} & \textbf{LS} &	\textbf{LO}	& \textbf{WS} & \textbf{WO} \\
\hline
\textbf{Total cost / k\$}	&	19.95 &	16.90 &	16.90	&	11.32	&	9.08 \\
\hline
\end{tabular}
\end{table}
}

\begin{figure}[t]
\centering
\includegraphics[width=0.45\textwidth]{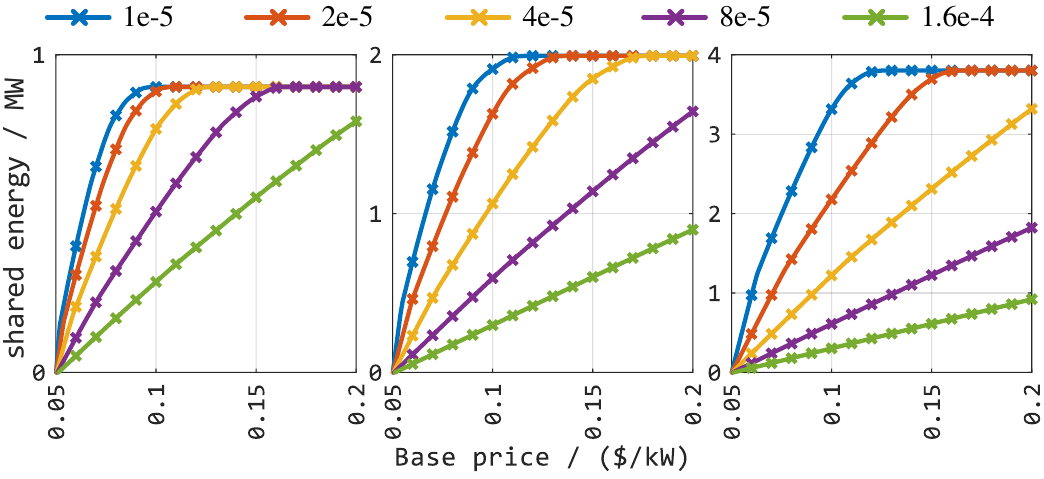}
\caption{Bidding curve $y_i \left( \varpi_i^0 \right)$ of LAMs with 50, 100, and 200 prosumers with different elasticities $a_i$.}
\label{fig_diff_a}
\end{figure}

The bidding curve $y_i \left( \varpi_i^0 \right)$ of LAMs with 50, 100, and 200 prosumers with different elasticities $a_i$ is shown in Fig. \ref{fig_diff_a}. A large $a_i$ may lead to the saturation phenomenon, while a small $a_i$ would decrease the price sensitivity of prosumers. Therefore, a moderate elasticity is necessary for energy sharing. Fig. \ref{fig_diff_a} shows that the range of moderate elasticity is approximately inverse proportion to the scale of the LAM. Hence, the value of elasticity is supposed to reduce as the market scale increases. It indicates that the market clearing result will approximate to the social optimum in large-scale markets as discussed in Remark \ref{remark_a}. Besides, Fig. \ref{fig_diff_a} also verifies the monotonicity of $y_i \left( \varpi_i^0 \right)$ disclosed in Proposition \ref{prop_monotonicity}.

\subsection{Sharing Price Without Electric Utility}
\begin{figure}[t]
\centering
\includegraphics[width=0.45\textwidth]{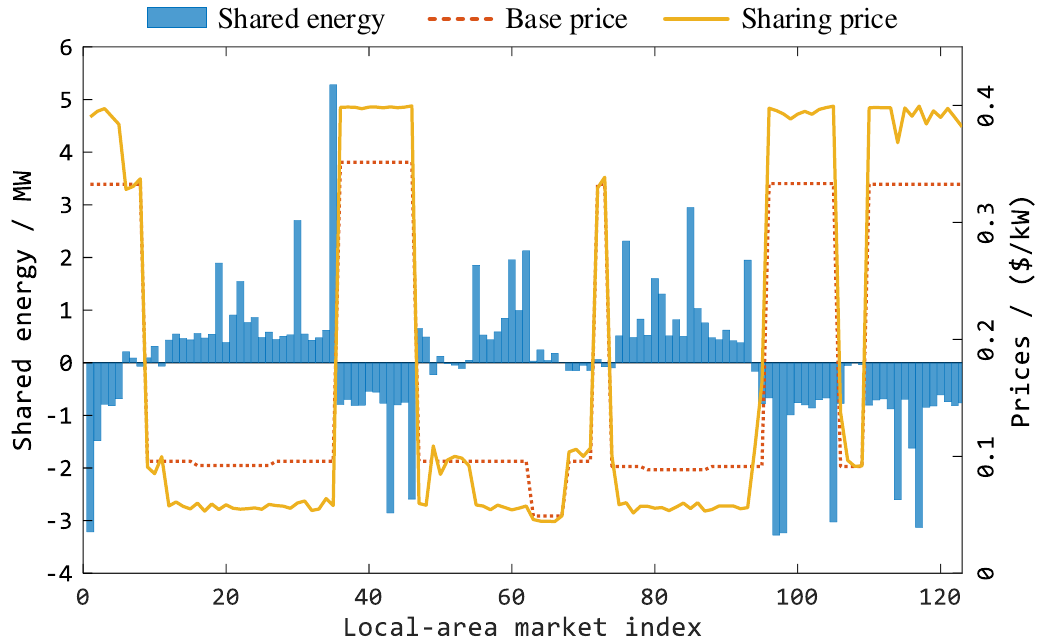}
\caption{Upper-level wide-area sharing market outcome without the electric utility.}
\label{fig_market_clear_noUtility}
\end{figure}

The WAM outcome without the electric utility is shown in Fig. \ref{fig_market_clear_noUtility}. It means that prosumers cannot trade with the electric utility, i.e., $p_j^+=p_j^-=0$. Without the competition of the electric utility, the energy sharing price is out of control. For the LAMs that purchases electricity through the WAM, the sharing prices sharply increase to almost 0.4 \$/kW, nearly twice of the buying price from the electric utility $\varpi^+ = 0.2$ \$/kW. Due to the line congestion, the sharing prices of LAMs 63-67 drop to lower than the selling price to the electric utility $\varpi^+ = 0.05$ \$/kW. Though the shared energy in the WAM changes little, the sharing price fluctuates more intensely than that in Fig. \ref{fig_market_clear}. It verifies that the electric utility takes effect in limiting the market power exploitation in the sharing market.

\section{Conclusion}
Wide-area sharing market can further promote the social efficient than local-area ones by dealing with the uneven distribution of energy resources in local communities. This paper proposes a two-layer energy sharing market with an upper-level wide-area market and numerous lower-level local-area markets. The local-area markets for prosumers in the same communities do not have to clear. The uncleared energy of a local-area market is shared with other local-area markets in the wide-area market. With the wide-area sharing, the social efficiency is further promoted. It is proved that the outcome of the proposed two-layer market approaches to the social optimum. Furthermore, to obtain the market outcome, this paper proposes an efficient hierarchically distributed bidding algorithm. Numerical experiments carried out on the IEEE 123-bus system with 11250 prosumers demonstrate the social efficiency of the two-layer sharing market and the effectiveness of the hierarchically distributed bidding algorithm. It is expected that this study can inspire more studies on wide-area energy sharing for large-scale distribution systems.

\bibliographystyle{IEEEtran}
\bibliography{mybib}

\begin{appendices}
\section{Proof of Proposition \ref{prop_complementary}} \label{appendix_complementary}
\setcounter{equation}{0}
\renewcommand\theequation{A.\arabic{equation}}

\begin{proof}
Given $x_j, \forall j\in \mathcal{U}_i \setminus \left\{j\right\}$, assume that there is an optimal solution to \eqref{prosumer_model} $s_j^* := \left\{p_j^*, p_j^{+*}, p_j^{-*}, x_j^*\right\}$ such that $p_j^{+*} > 0, p_j^{-*} > 0$. Then we can construct a strictly smaller solution than $s_j^*$.

Define $p_{j,min} = \min \left\{ p_j^{+*}, p_j^{-*} \right\} > 0$. Construct a new solution $s'_j := \left\{p_j^*, p_j^{+*} - p_{j,min}, p_j^{-*} - p_{j,min}, x_j^*\right\}$. It is easy to check that $s'_j$ is feasible and that the purchased and sold electricity are complementary. Meanwhile, the objective value is smaller than the optimal value by $(\varpi^+ - \varpi^-) p_{j,min} > 0$, which completes the proof by contradiction.
\end{proof}

\section{Proof of Proposition \ref{prop_NE_equivalent_solution}} \label{appendix_NE_equivalent_solution}
\setcounter{equation}{0}
\renewcommand\theequation{B.\arabic{equation}}

\begin{proof}
We first prove that the NE is the optimal solution to problem \eqref{equivalent_problem} and then prove its existence and uniqueness.

1) \textit{Equivalent transformation.} Suppose there is an NE $\bm{s}^{i*} := \left\{ p_j^*, p_j^{+*}, p_j^{-*}, x_j^*, \forall j\in\mathcal{U}_i \right\}$. Since the NE is the optimal solution to \eqref{Nash_game}, it satisfies the optimality condition \cite[Thm. 3.24]{ruszczynski2006nonlinear}
\begin{align} \label{KKT1}
\left< \nabla \mathcal{F}_j (s_j^*, \bm{s}_{-j}^*), s_j - s_j^* \right> \ge 0, ~\forall s_j\in\mathcal{S}_j, j \in \mathcal{U}_i,
\end{align}
where the gradient $\nabla \mathcal{F}_j (s_j^*, \bm{s}_{-j}^*)$ is defined as
\begin{align}
\nabla \mathcal{F}_j (s_j^*, \bm{s}_{-j}^*) = 
\left[ \begin{matrix}
c_j p_j^* + b_j \\
\varpi^+ \\
-\varpi^- \\
a_i x_j^* - \varpi_i^0 + a_i \sum_{j'\in\mathcal{U}_i} x_{j'}^*
\end{matrix} \right]
\end{align}

Note that the feasibility region of problem \eqref{equivalent_problem} is the Cartesian product of all action sets $\mathcal{S}_j, \forall j\in \mathcal{U}_i$. It directly induces that the optimality condition of \eqref{equivalent_problem} is the collection of inequalities \eqref{KKT1} for all $j\in \mathcal{U}_i$. Therefore, the NE of \eqref{Nash_game} is equivalent to the optimal solution to \eqref{equivalent_problem}.

2) \textit{Existence of optimal solution.} Problem \eqref{equivalent_problem} is obviously feasible. If the objective value is always bounded, there must exist an optimal solution. From the weak duality \cite[Sec. 5.2.2]{boyd2004convex}, the objective value of an arbitrary feasible solution to the dual problem provides a lower bound to the primal problem. Therefore, the existence of an optimal solution to problem \eqref{equivalent_problem} is equivalent to the feasibility of its dual problem. The constraints of the dual problem are given by
\begin{align}
\varpi^+ - \mu_j^{+} - \mu_j^{PB} =~ & 0, ~\forall j \in \mathcal{U}_i \\
-\varpi^- - \mu_j^{-} + \mu_j^{PB} =~ & 0, ~\forall j \in \mathcal{U}_i \\
\underline{\mu}_j, \overline{\mu}_j, \mu_j^{+}, \mu_j^{-}, \mu_j^{PB} \ge~ & 0, ~\forall j \in \mathcal{U}_i.
\end{align}

We can construct a feasible solution as $\underline{\mu}_j = \overline{\mu}_j = \mu_j^{+} = 0$, $\mu_j^{-} = \varpi^+ - \varpi^-$, $\mu_j^{PB} = \varpi^+, ~\forall j \in \mathcal{U}_i$, which proves the existence of optimal solution.

3) \textit{Uniqueness of optimal solution.} Assume that there are two different optimal solutions to problem \eqref{equivalent_problem} denoted by $\widehat{\bm{s}}^i := \{ \widehat{p}_j, \widehat{p}_j^+, \widehat{p}_j^-, \widehat{x}_j, \forall j\in\mathcal{U}_i \}$ and $\check{\bm{s}}^i := \{ \check{p}_j, \check{p}_j^+, \check{p}_j^-, \check{x}_j, \forall j\in\mathcal{U}_i \}$. Suppose that the two solutions differ in some decisions of prosumers indexed by $j\in \mathcal{U}^d \neq \emptyset$. By Proposition \ref{prop_complementary}, at the NE, given $\left\{p_j, x_j\right\}$, there exist a unique and defined $\left\{p_j^+, p_j^-\right\}$. Thus for each prosumer $j\in \mathcal{U}^d$, their $\left\{p_j, x_j\right\}$ must be not identical, i.e., $\left\{ \widehat{p}_j, \widehat{x}_j \right\} \neq \left\{ \check{p}_j, \check{x}_j \right\}, \forall j\in \mathcal{U}^d$.

Divide the objective of problem \eqref{equivalent_problem} into two parts: 1) a strongly convex function $\mathcal{F}^{(1)} (\bm{p}^i, \bm{x}^i)$ and 2) a linear function $\mathcal{F}^{(2)} (\bm{p}^{i+}, \bm{p}^{i-})$. Let $\mathcal{F}^*$ denote the optimal value of problem \eqref{equivalent_problem}, i.e., $\mathcal{F}^* = \mathcal{F}^{(1)} (\widehat{\bm{p}}^i, \widehat{\bm{x}}^i) + \mathcal{F}^{(2)} (\widehat{\bm{p}}^{i+}, \widehat{\bm{p}}^{i-}) = \mathcal{F}^{(1)} (\check{\bm{p}}^i, \check{\bm{x}}^i) + \mathcal{F}^{(2)} (\check{\bm{p}}^{i+}, \check{\bm{p}}^{i-})$.

We can construct another solution $\widetilde{s}^i = (\widehat{\bm{s}}^i + \check{\bm{s}}^i) / 2$, which is obviously feasible. Since the optimal solutions $\widehat{\bm{s}}^i$ and $\check{\bm{s}}^i$ differ in $\left\{\bm{p}^i, \bm{x}^i\right\}$, by the strong convexity of $\mathcal{F}^{(1)} (\bm{p}^i, \bm{x}^i)$ and the linearity of $\mathcal{F}^{(2)} (\bm{p}^{i+}, \bm{p}^{i-})$, we have
\begin{align}
\mathcal{F}^{(1)} (\widetilde{\bm{p}}^i, \widetilde{\bm{x}}^i) + \mathcal{F}^{(2)} & (\widetilde{\bm{p}}^{i+}, \widetilde{\bm{p}}^{i-}) \notag \\
<& (\mathcal{F}^{(1)} (\widehat{\bm{p}}^i, \widehat{\bm{x}}^i) + \mathcal{F}^{(1)} (\check{\bm{p}}^i, \check{\bm{x}}^i)) / 2 \notag \\
&+ (\mathcal{F}^{(2)} (\widehat{\bm{p}}^{i+}, \widehat{\bm{p}}^{i-}) + \mathcal{F}^{(2)} (\check{\bm{p}}^{i+}, \check{\bm{p}}^{i-})) / 2 \notag \\
=& \mathcal{F}^*
\end{align}
It means that if there are non-unique optimal solutions to problem \eqref{equivalent_problem}, we can always find another solution with a smaller objective value than the optimal value. The uniqueness of the optimal solution is proved by contradiction, which completes the proof.
\end{proof}

\section{Proof of Proposition \ref{prop_pricing_relationship}} \label{appendix_pricing_relationship}
\setcounter{equation}{0}
\renewcommand\theequation{C.\arabic{equation}}

\begin{proof}
For each prosumer $j\in\mathcal{U}_i$, given $\bm{x}_{-j}^*$, the NE satisfies the KKT condition of problem \eqref{prosumer_model}
\begin{align}
& c_j p_j^* + b_j - \underline{\mu}_j^* + \overline{\mu}_j^* - \mu_j^{PB*} = 0 \label{KKT_p} \\
& \varpi^+ - \mu_j^{+*} - \mu_j^{PB*} = 0 \label{KKT_p_plus} \\
& -\varpi^- - \mu_j^{-*} + \mu_j^{PB*} = 0 \label{KKT_p_minus} \\
& -\varpi_i^0 + a_i x_j^* + a_i \sum_{j'\in\mathcal{U}_i} x_{j'}^* + \mu_j^{PB*} = 0 \label{KKT_x} \\
& 0 \le p_j^* - \underline{P}_j \perp \underline{\mu}_j^* \ge 0 \label{KKT_mu_p1} \\
& 0 \le \overline{P}_j - p_j^* \perp \overline{\mu}_j^* \ge 0 \label{KKT_mu_p2} \\
& 0 \le p_j^{+*} \perp \mu_j^{+*} \ge 0 \label{KKT_mu_p_plus} \\
& 0 \le p_j^{-*} \perp \mu_j^{-*} \ge 0 \label{KKT_mu_p_minus} \\
& D_j + x_j^* + p_j^{-*} = p_j^* + p_j^{+*} \label{KKT_pb}
\end{align}

a) Combining \eqref{KKT_p_plus} and \eqref{KKT_mu_p_plus}, we have
\begin{align*}
\varpi^+ - \mu_j^{PB*} = \mu_j^{+*} \ge 0.
\end{align*}
Similarly, the combination of \eqref{KKT_p_minus} and \eqref{KKT_mu_p_minus} yields
\begin{align*}
\mu_j^{PB*} - \varpi^- = \mu_j^{-*} \ge 0.
\end{align*}

b) Recalling the definition of $\varpi_i ( \bm{x}^{i*} )$, equation \eqref{KKT_x} directly yields
\begin{align*}
x_j^* = \frac{\varpi_i ( \bm{x}^{i*} ) - \mu_j^{PB*}}{a_i}.
\end{align*}

c) Summing \eqref{KKT_x} over all $j\in\mathcal{U}_i$, we have
\begin{align*}
-|\mathcal{U}_i|\varpi_i^0 + a_i (1 + |\mathcal{U}_i|) \sum_{j\in\mathcal{U}_i} x_j^* + \sum_{j\in\mathcal{U}_i} \mu_j^{PB*} = 0
\end{align*}
By the definition of $\varpi_i ( \bm{x}^{i*} )$, equation \eqref{sharing_price_equation} holds directly.

d) By \eqref{sharing_price_equation}, the value of the energy sharing price $\varpi_i ( \bm{x}^{i*} )$ is the average of the base price $\varpi_i^0$ and shadow prices of prosumers. Recall that $\mu_j^{PB*}, \forall j\in\mathcal{U}_i$ are in the range $\left[\varpi^-, \varpi^+\right]$. If $\varpi_i^0$ is also bounded by $\varpi^+$ and $\varpi^-$, their average is obviously located in the same range.

As the scale of prosumers increases, we have
\begin{align*}
\lim_{|\mathcal{U}_i| \to \infty} \varpi_i ( \bm{x}^{i*} ) &= \lim_{|\mathcal{U}_i| \to \infty} \frac{\varpi_i^0 + \sum_{j\in \mathcal{U}_i} \mu_j^{PB*}}{1 + |\mathcal{U}_i|} \\
&= \lim_{|\mathcal{U}_i| \to \infty} \frac{\sum_{j\in \mathcal{U}_i} \mu_j^{PB*}}{|\mathcal{U}_i|}.
\end{align*}
Even if $\varpi_i^0$ is out of $\left[\varpi^-, \varpi^+\right]$, the energy sharing price also approaches the range, which completes the proof.
\end{proof}

\section{Proof of Proposition \ref{prop_monotonicity}} \label{appendix_monotonicity}
\setcounter{equation}{0}
\renewcommand\theequation{D.\arabic{equation}}

\begin{proof}
We will prove it by contradiction. Assume that $y_i \left( \varpi_i^0 \right)$ is monotonically decreasing with respect to $\varpi_i^0$. At the NE, a larger $\varpi_i^0$ causes a smaller $y_i \left( \varpi_i^0 \right)$. Then by the definition \eqref{CESM_pricing}, the energy sharing price $\varpi_i ( \bm{x}^{i*} )$ increases. Then discuss by the trend of $\mu_j^{PB*}$.

a) \textbf{$\mu_j^{PB*}$ decreases.} By \eqref{KKT_p}, \eqref{KKT_mu_p1} and \eqref{KKT_mu_p2}, the power generation is given by
\begin{align}
p_j^* = \left[ \frac{\mu_j^{PB*} - b_j}{c_j} \right]_{\overline{P}_j}^{\underline{P}_j},
\end{align}
where $\left[\cdot\right]_a^b$ is the projection onto the range $\left[a, b\right]$. Thus $p_j^*$ decreases or remains. By \eqref{KKT_p_plus} and \eqref{KKT_mu_p_plus}, if $\mu_j^{PB*} \in \left[\varpi^-, \varpi^+\right)$, we have $\mu_j^{+*} > 0$, which indicates $p_j^{+*} = 0$. Only if $\mu_j^{PB*} = \varpi^+$, $p_j^{+*}$ may be positive. Therefore, in this situation, $p_j^{+*}$ decreases or remains zero. Similarly, $p_j^{-*}$ may be positive if $\mu_j^{-*} = \varpi^-$; otherwise, it remains zero. Thus $p_j^{-*}$ increases or remains zero in this situation. Besides, \eqref{KKT_x} indicates that $x_j^*$ must increase. However, the trends of $p_j^*, p_j^{+*}, p_j^{-*}$ and $x_j^*$ do not satisfy the power balance constraint \eqref{KKT_pb}. It means that the situation of decreasing $\mu_j^{PB*}$ is impossible.

b) \textbf{$\mu_j^{PB*}$ increases.} Similarly to a), an increasing $\mu_j^{PB*}$ can deduce that $p_j^*$ and $p_j^{+*}$ increase or remain, while $p_j^{-*}$ decreases or remains zero. By the power balance constraint \eqref{KKT_pb}, $x_j^*$ will decrease or remain, which does not violate \eqref{KKT_x}. Therefore, in the situation of increasing $\mu_j^{PB*}$, $x_j^*$ will increase or remain.

c) \textbf{$\mu_j^{PB*}$ remains.} By \eqref{KKT_x}, $x_j^*$ must increase. There is an possible condition, where $\mu_j^{PB*} \equiv \varpi^+$ and $p_j^{+*}$ increases, while $p_j^*$ and $p_j^{-*}$ remain.

To sum up, except for the impossible situation of decreasing $\mu_j^{PB*}$, the other two situations can deduce that $x_j^*$ will increase or remain. Summing over $\forall j\in\mathcal{U}_i$, $y_i \left( \varpi_i^0 \right) = \sum_{j\in\mathcal{U}_i} x_j^*$ also increases or remains. It does not match the initial assumption of decreasing $y_i \left( \varpi_i^0 \right)$, which proves the original proposition by contradiction.
\end{proof}

\section{Proof of Proposition \ref{prop_MPEC}} \label{appendix_MPEC}
\setcounter{equation}{0}
\renewcommand\theequation{E.\arabic{equation}}

\begin{proof}
We first prove that Algorithm \ref{algo_DESM_bidding} is equivalent to the dual decomposition method to solve the problem \eqref{equivalent_problem_total}. Following \cite{su2023hierarchically, palomar2006tutorial}, define the Lagrangian of problem \eqref{equivalent_problem_total}
\begin{align}
L = & \sum_{i\in\mathcal{N}} \sum_{j\in\mathcal{U}_i} \Big\{ \frac{c_j}{2} p_j^2 + b_j p_j + \varpi^+ p_j^+ - \varpi^- p_j^- \Big\}  \notag \\
& + \sum_{i\in\mathcal{N}} \Big\{ \frac{a_i}{2} y_i^2 + \frac{a_i}{2} \sum_{j\in\mathcal{U}_i} x_j^2 \Big\} \notag \\
& + \lambda^{PB} \sum_{i\in\mathcal{N}} y_i + \sum_{l\in\mathcal{L}} \lambda_l \Big( \sum_{i\in\mathcal{N}} \pi_{il} ~ y_i - F_l \Big)
\end{align}
where $\lambda^{PB}$ and $\lambda_l$ are the Lagrangian multiplier of \eqref{cons_power_balance} and \eqref{cons_power_flow}, respectively.

The dual decomposition method is alternately solving subproblems for $\forall i \in \mathcal{N}$
\begin{subequations} \label{DD_sub}
\begin{align}
\min & \sum_{j\in\mathcal{U}_i} \Big\{ \frac{c_j}{2} p_j^2 + b_j p_j + \varpi^+ p_j^+ - \varpi^- p_j^- \Big\} + \frac{a_i}{2} \sum_{j\in\mathcal{U}_i} x_j^2 \notag \\
& + \frac{a_i}{2} y_i^2 + (\lambda^{PB} (k) + \sum_{l\in\mathcal{L}} \pi_{il} \lambda_l(k)) y_i \\
\text{s.t.} ~~ & \left\{ p_j, p_j^+, p_j^-, x_j \right\} \in \mathcal{S}_j, \forall j\in\mathcal{U}_i \\
& y_i = \sum_{j\in \mathcal{U}_i} x_j
\end{align}
\end{subequations}
to attain $y_i (k), \forall i \in \mathcal{N}$ and update Lagrangian multipliers
\begin{subequations} \label{DD_master}
\begin{align}
&\lambda^{PB} (k+1) = \lambda^{PB} (k) + \alpha^{PB} \sum_{i\in\mathcal{N}} y_i (k) \\
&\lambda_l (k+1) = \left[ \lambda_{l} (k) + \alpha_l \left( \sum_{i\in\mathcal{N}} \pi_{il} ~ y_i (k) - F_l \right) \right]^+
\end{align}
\end{subequations}
where $\left[\cdot\right]^+$ is the projection onto the range $\left[0, +\infty \right)$.

Let $\lambda^{PB} \equiv - \varpi^{PB}$ and $\lambda_l \equiv - \varpi_l$. Then updating Lagrangian multipliers \eqref{DD_master} is equivalent to adjusting prices \eqref{DESM_bidding}. Meanwhile, solving subproblem \eqref{DD_sub} equals to seeking the NE of the LAM according to Proposition \ref{prop_NE_equivalent_solution}. Therefore, Algorithm \ref{algo_DESM_bidding} converges to the optimal solution to problem \eqref{equivalent_problem_total}.

The existence and uniqueness of the optimal solution to problem \eqref{equivalent_problem_total} is just the same as that of problem \eqref{equivalent_problem} in Appendix \ref{prop_NE_equivalent_solution}. Due to the
space limitation, the proof is omitted here.
\end{proof}

\end{appendices}
\end{document}